\documentclass{article}

\usepackage[final]{neurips_2022}

\usepackage[utf8]{inputenc} 
\usepackage[T1]{fontenc}    
\usepackage{hyperref}       
\usepackage{url}            
\usepackage{booktabs}       
\usepackage{amsfonts}       
\usepackage{nicefrac}       
\usepackage{microtype}      
\usepackage{xcolor}         
\usepackage{graphicx}

\usepackage{todonotes}

\title{Towards Openness Beyond Open Access: User Journeys through 3 Open AI Collaboratives}

\author{%
  Jennifer Ding \\
  The Alan Turing Institute \\
  \texttt{jding@turing.ac.uk} \\
  \And
  Christopher Akiki \\
  Leipzig University \\
  \texttt{christopher.akiki@uni-leipzig.de} \\
  \AND
  Yacine Jernite \\
  Hugging Face \\
  \texttt{yacine@huggingface.co} \\
  \And
  Anne Lee Steele \\
  The Alan Turing Institute \\
  \texttt{asteele@turing.ac.uk} \\
  \And
  Temi Popo \\
  Mozilla Foundation \\
  \texttt{temi@mozillafoundation.org} \\
}
\begin{document}
\maketitle
\begin{abstract}
Open Artificial Intelligence (Open source AI) collaboratives offer alternative pathways for how AI can be developed beyond well-resourced technology companies and who can be a part of the process. To understand how and why they work and what additionality they bring to the landscape, we focus on three such communities, each focused on a different kind of activity around AI: building \emph{models} (BigScience workshop), \emph{tools/ways of working} (The Turing Way), and \emph{ecosystems} (Mozilla Festival’s Building Trustworthy AI Working Group). First, we document the community structures that facilitate these distributed, volunteer-led teams, comparing the collaboration styles that drive each group towards their specific goals. Through interviews with community leaders, we map user journeys for how members discover, join, contribute, and participate. Ultimately, this paper aims to highlight the diversity of AI work and workers that have come forth through these collaborations and how they offer a broader practice of openness to the AI space.
\end{abstract}

\section{Introduction}
\label{sec:introduction}
While the majority of AI production and resources are concentrated within technology companies in the US, Europe, and China~\citep{savage2020race}, the growth of open AI collaboratives offer alternative pathways for how AI is developed and who is able to be a part of the process. In addition to creating open-access resources, these online, distributed, and largely volunteer-led collaboratives create new opportunities for more people from outside of the technology field to participate in the process of building, deploying, and governing AI. This kind of environment enables more actors and activities to become open for a broader practice of open AI.

This paper highlights three open AI communities each focused on a different kind of activity around AI---building \emph{models} (\href{https://bigscience.huggingface.co}{BigScience Workshop}), \emph{tools/ways of working} (\emph{The Turing Way})~\citep{the_turing_way_community_2022_6909298}, and \emph{ecosystems} (\href{https://www.mozillafestival.org/en/working-groups/}{Mozilla Festival’s Trustworthy AI Working Group})---and typical user journeys taken by their members to discover, join, contribute, and lead within the team. Though there are other such communities, these three were chosen due to the availability of open materials (e.g. meeting notes or project documentation via platforms like GitHub and Hugging Face Hub) and regular meetings open to the public. In addition to referencing public community materials, we have conducted qualitative interviews with community leaders to understand explicit and implicit structures that influence member experience.

\begin{figure}[ht!]
\centering
\includegraphics[width=0.8\textwidth]{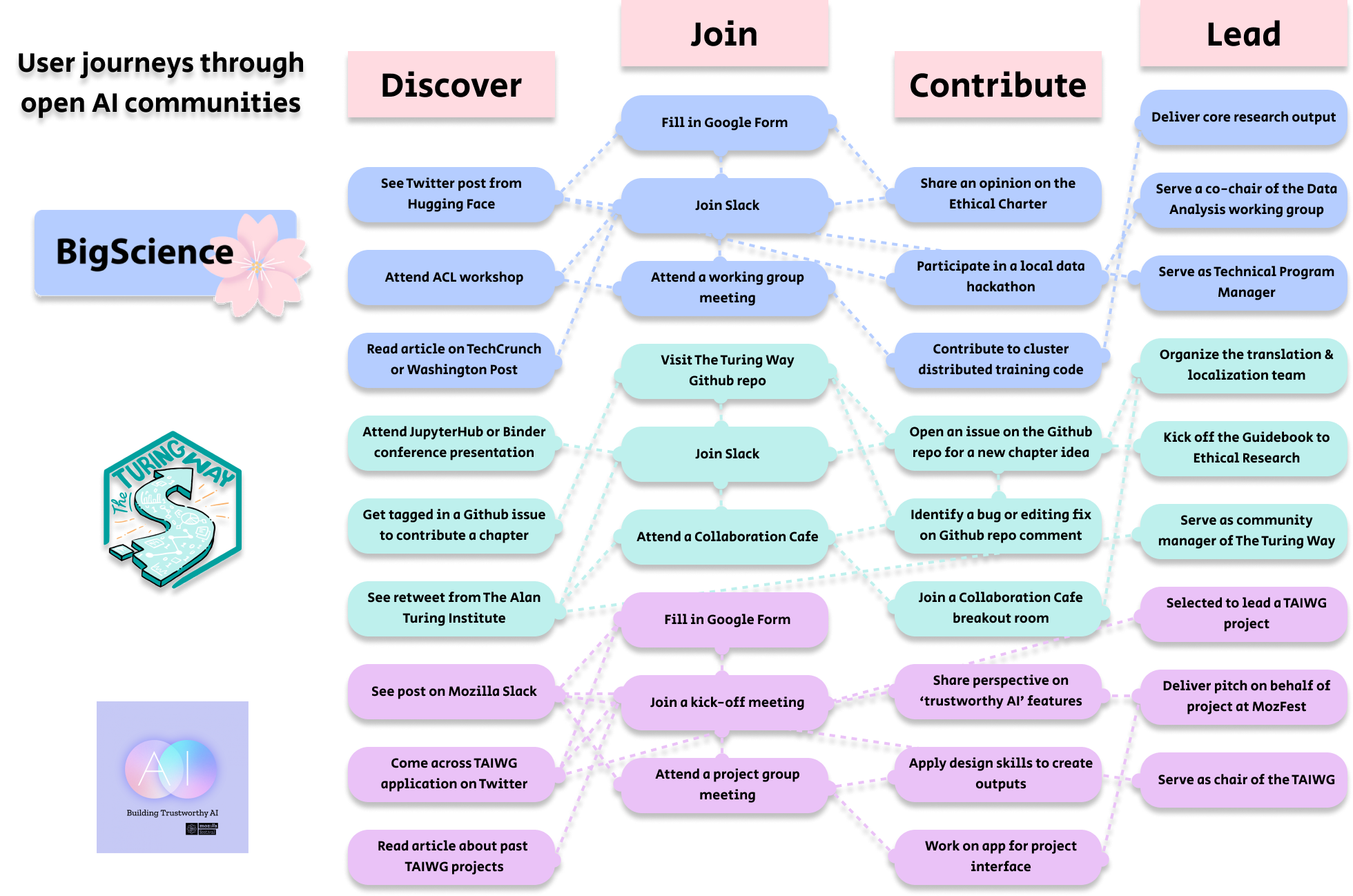}
\caption{User journeys through open AI communities}
\label{fig:userjourneys}
\end{figure}

\section{Community Structures}
\label{sec:community-structures}
\begin{figure}[ht!]
\centering
\includegraphics[width=0.85\textwidth]{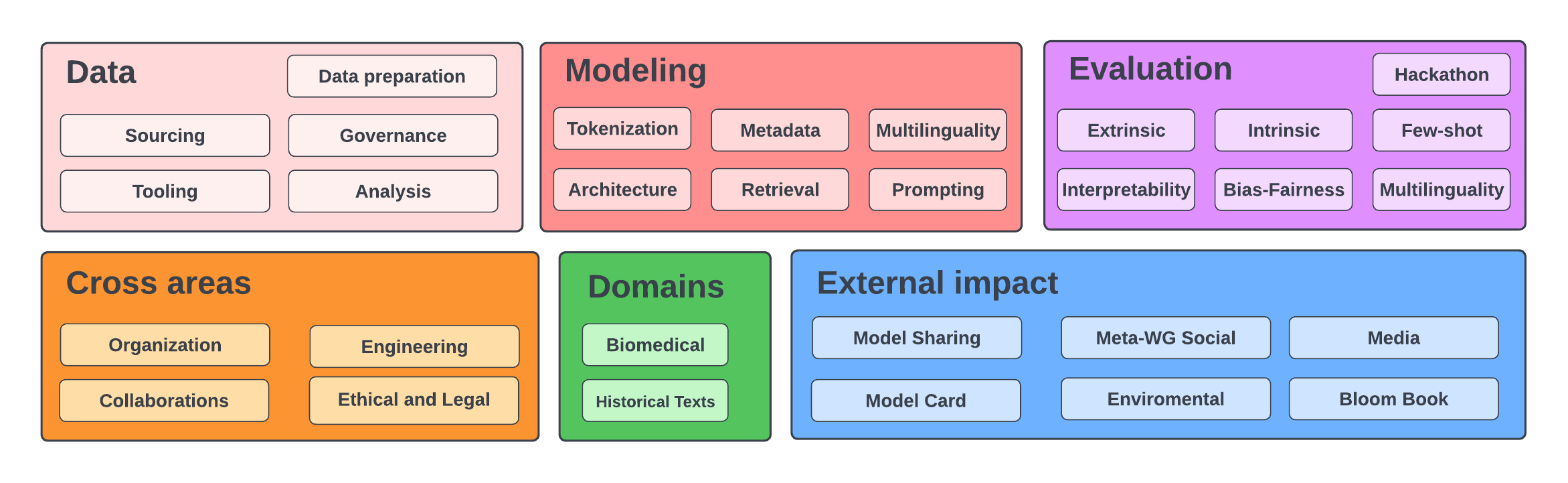}
\caption{Division of BigScience Workshop into working groups}
\label{fig:workinggroups}
\end{figure}

\paragraph{BigScience Workshop} The BigScience Workshop was a value-driven~\citep{elliott-tapestry-values} research initiative modeled after large-scale collaboration schemes from the second half of the twentieth century~\citep{sep-longino-scientific-knowledge-social} to address research challenges in particle physics, genetics, and astronomy by convening large groups of researchers organized in specialized subgroups and instrumentalizing specialized hardware. Inspired by these initiatives, BigScience Workshop assembled over 1000~volunteer researchers from May~2021 to July~2022, to work together toward training the BLOOM (BigScience Large Open-science Open-access Multilingual) Language Model. The workshop was composed of working groups focusing on topics like multilinguality, evaluation of bias-fairness, data governance, and environmental impact (see: Figure~\ref{fig:workinggroups}). Though the workshop has ended, members continue to collaborate, though with less intensity than before. Through GitHub contributor records, we find that members come from countries such as France, US, India, Saudi Arabia, Indonesia, Germany and Singapore.  

\begin{figure}[t!]
\centering
\includegraphics[width=\textwidth]{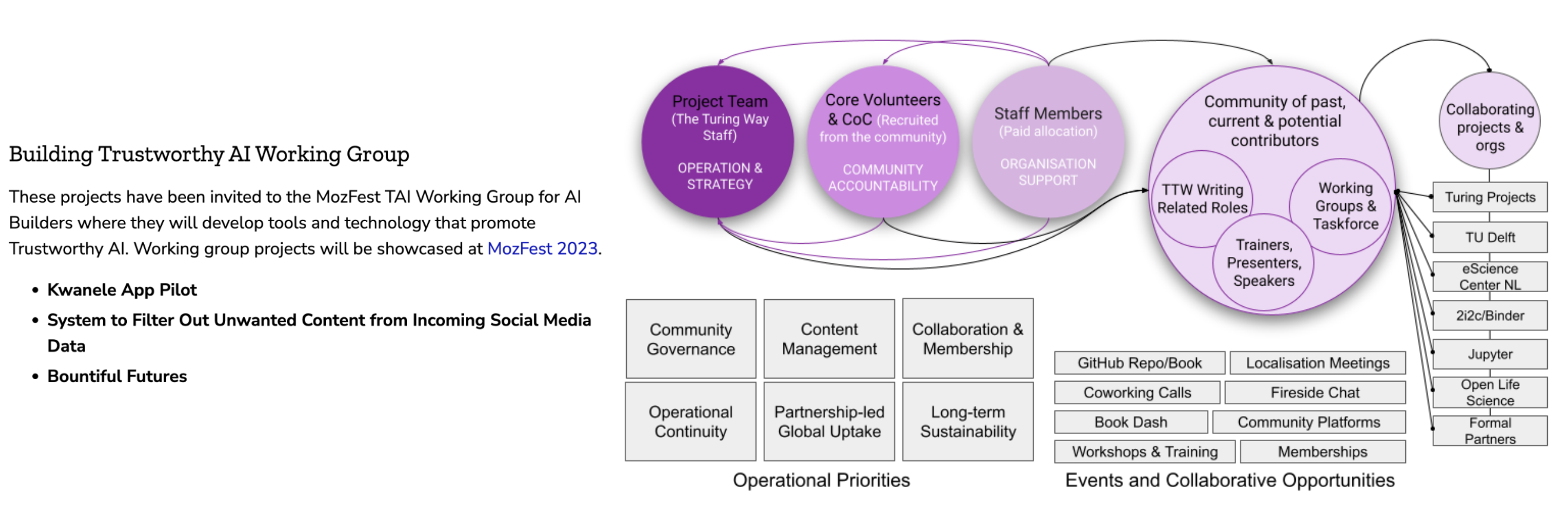}
\caption{Left: MozFest Trustworthy AI Working Group Cohort 3, Right: \emph{The Turing Way} Roles}
\label{fig:taiwgttw}
\end{figure}

\paragraph{The Turing Way} Created in 2019, \emph{The Turing Way} is a distributed community of researchers and practitioners from data-science related fields who are co-creating a handbook of tools and best practices to ensure that conducting open, responsible, localised, and collaborative data science is "too easy not to do.” The book is co-written by over 400~volunteers in multiple languages through GitHub, which serves as a data store, text version control, and an asynchronous collaboration tool (see: Figure~\ref{fig:taiwgttw}). The community convenes through i)~bi-weekly Collaboration Cafes for co-working and ii)~biannual Book Dash events for community strategizing, as well as contributing to and maintaining the repo. \emph{The Turing Way} handbook is composed of five guidebooks for Reproducible Research, Project Design, Communication, Collaboration, and Ethical Research. Through GitHub contributor records, we find that members come from countries such as the UK, The Netherlands, India, Saudi Arabia, Argentina, and the US~\citep{the_turing_way_community_2022_6909298}.

\paragraph{Mozilla Festival's Trustworthy AI Working Group} As part of the Mozilla community, the Building Trustworthy AI Working Group is composed of over 400 global members who collaborate on projects selected by the core leadership team. The TAIWG is led by a chair and members include project leads and volunteers who join projects they are interested in (see: Figure~\ref{fig:taiwgttw}). The TAIWG began in 2020 and it runs on an annual schedule, with cohorts of projects kicking off in the Fall to work towards the Mozilla Festival in the Spring. Over 20 projects have graduated so far, and \href{https://wiki.mozilla.org/Working_Groups}{past projects} include MOSafely: An AI Community Making the Internet a Safer Place for Youth, a feminist dictionary in AI and AI Governance in Africa. On the \href{https://foundation.mozilla.org/en/blog/youre-invited-to-build-trustworthy-ai-with-mozfest/}{working group page}, we see that members come from countries like South Africa, Canada, US, and the UK.

\section{User Journeys}
\label{sec:user-journeus}
To understand the range of member experiences within the three open AI communities, we have conducted interviews with community leaders to map user journeys for key points in a member’s experience with a community: discovering, joining, contributing, and leading.

\paragraph{Discovering} Engaging with a community begins with discovery, a growing challenge in AI where new initiatives are emerging rapidly. For all three communities, the influence of prominent members and the reputation of organizations supporting the groups (e.g. HuggingFace, The Alan Turing Institute, and Mozilla Foundation) played important roles. However, in order to expand reach beyond existing networks, the communities applied tactics to diversify their membership. BigScience’s founding team took steps to outreach outside their geographical and professional domains through situated events like local data hackathons. \emph{The Turing Way} seeks out collaborations with open science organizations around the world to build off each other’s work and connect their communities. The MozFest TAIWG is made up of engaged members from the Mozilla Festival Community and the Mozilla Developer Community. To expand our reach, they have also invited participation from local partners in cities where the Mozilla Festival is held.

A common draw for new members is in a shared challenge and space for addressing it. Whether it’s an openly developed LLM, improving research culture, or building trustworthy AI, the communities center on a direction of AI work that may not be accessible elsewhere for aspiring members.

\paragraph{Joining} Joining is an important point in a user journey that can be defined by the barriers presented. A comparison can be made to the process of joining an academic or industry AI research lab, where an individual must undergo years of accreditation and a gauntlet of interviews to even be considered for entry. This process is a major limiting factor, and as AI becomes more impactful in everyday life, this barrier to entry exacerbates the power dynamic between AI producers and everyone else. 

In contrast, each of the three collaboratives is open to any interested participant who typically find out about the community and join through digital doorways such as an online channel (e.g. email listserv, Twitter), collaboration space (e.g. Slack), or video call meeting or event (e.g. Zoom). Because these channels are available to anyone with access to these online resources, this means many more people in different time zones, backgrounds, and skill levels can enter.

While lowering the barrier to entry is the first step to the joining process, it is not enough to facilitate active participation. In our conversations with community leaders, they shared that the act of “lurking” is common practice and not something to be stigmatized \cite{chenchanglurk}. Whether it’s listening in on meetings or consuming and reacting to content on Slack, this behavior is characteristic of digital spaces and offers an easy, safe way to explore the community before contributing.

\paragraph{Contributing} After a member has passed through the digital doorway to join a community, the next step in their user journey is to begin contributing. Community leaders remarked on how the kinds of available "jumping in" points varied by the stage of the community. This was particularly important for BigScience and the TAIWG which have hard deadlines associated with working groups, which members join based on their skill sets and interests. Because the core activity for \emph{The Turing Way} is co-writing, members can easily join at any point in time. However, the process of contributing via GitHub may not be straightforward to new members. Thus, first contributions often require a new member to pair with older members who guide them through the process. 

Because contributing to all three communities is voluntary, the question arises for what motivates members to contribute. Community leaders shared that members are often driven by an interest in working on a problem that they are not able or empowered to do in their normal jobs, to gain skills and experience in AI, and capture their work in papers, blog posts, tools, or presentations. However, they also shared that identifying opportunities to “give back” to members is important, whether through awards/recognition or connecting them to other opportunities in the community and beyond.

\paragraph{Leading} Leadership takes on different forms within all three communities. In addition to more traditional leadership roles such as leading a working group, program and community management were also leadership roles that the communities invested in. The \href{https://noidea.dog/glue}{glue work} associated with these latter roles is important for many collaborations, and crucial for these collaboratives where membership is composed of global volunteers.

Typical leadership pathways come from an invitation from an existing leader or through responsibility for a core work stream. In BigScience and \emph{The Turing Way}, there are leadership roles filled through a formal hiring process via HuggingFace and The Alan Turing Institute, respectively. Hired members serve as a core engine to drive the project forward during working hours, managing funding and logistics and providing infrastructure and resources. However, any member can, through expertise or initiative, propose a new project to lead and recruit collaborators to carry it through. 

Community leaders shared stories of moments where a member was formally or informally empowered to leadership. These ranged from small actions such as being asked to share their opinion in a meeting to longer processes where continuous work on a project organically led to a member’s implicit status as the de facto owner of it. Because all three communities draw people from a range of backgrounds, who may not see themselves as “technical” or “AI experts”, the communities offer a form of accreditation and empowerment through association with the group and recognition on research papers and through official titles (e.g. “AI Builder” in MozFest TAIWG).
\section{Towards a broader definition open AI}
\label{sec:broader-definition}
BigScience Workshop, \emph{The Turing Way}, and MozFest’s Building Trustworthy AI Working Group provide examples for how AI collaborations can diversify, democratize, and broaden our understanding of what open AI means. In addition to creating space for more people to join in and contribute to the AI field, they have also constructed environments where new ideas can emerge and new people are empowered to carry them out. Our community research shows that by lowering the barriers to entry through public, digital doorways and by creating space for new research directions, open AI collaboratives fill an important gap in the wider AI ecosystem. These three communities offer new frameworks to empower more people around the world to participate and shape the AI ecosystem in ways that are meaningful to them. Though some of these groups may disband in time, their example can serve as a template for future collaborations and to help accelerate a broader practice of open AI.

\begin{ack}
This work was supported by Towards Turing 2.0 under the EPSRC Grant EP/W037211/1 \& The Alan Turing Institute’. This work was supported by Wave 1 of The UKRI Strategic Priorities Fund under the EPSRC Grant EP/T001569/1 and EPSRC Grant EP/W006022/1, particularly the “Tools, Practices \& Systems” theme within those grants \& The Alan Turing Institute’.

The BigScience Workshop was granted access to the HPC resources of the Institut du développement et des ressources en informatique scientifique (IDRIS) du Centre national de la recherche scientifique (CNRS) under the allocation 2021-A0101012475 made by Grand équipement national de calcul intensif (GENCI). Model training ran on the Jean-Zay cluster of IDRIS, and we thank the IDRIS team for their responsive support throughout the project, in particular Rémi Lacroix.

This work was made possible by the Mozilla Foundation and its generous community, particularly the MozFest Trustworthy AI working group for AI Builders. We are thankful for the colleagues - past and present, community members, and project leads that help us build a healthier internet and more equitable automated future for all.
\end{ack}

\bibliographystyle{chicago}
\bibliography{bibliography}

\begin{thebibliography}{}

\bibitem[\protect\citeauthoryear{Chen and Chang}{Chen and
  Chang}{2013}]{chenchanglurk}
Chen, F.-C. and H.-M. Chang (2013).
\newblock Engaged lurking: The less visible form of participation in online
  small group learning.
\newblock {\em Research and Practice in Technology Enhanced Learning\/}~{\em
  8\/}(1), 171–199.

\bibitem[\protect\citeauthoryear{Elliott}{Elliott}{2017}]{elliott-tapestry-values}
Elliott, K.~C. (2017, 02).
\newblock {\em {A Tapestry of Values: An Introduction to Values in Science}}.
\newblock Oxford University Press.

\bibitem[\protect\citeauthoryear{Longino}{Longino}{2019}]{sep-longino-scientific-knowledge-social}
Longino, H. (2019).
\newblock {The Social Dimensions of Scientific Knowledge}.
\newblock In E.~N. Zalta (Ed.), {\em The {Stanford} Encyclopedia of
  Philosophy\/} ({S}ummer 2019 ed.). Metaphysics Research Lab, Stanford
  University.

\bibitem[\protect\citeauthoryear{Savage}{Savage}{2020}]{savage2020race}
Savage, N. (2020).
\newblock The race to the top among the world's leaders in artificial
  intelligence.
\newblock {\em Nature\/}~{\em 588\/}(7837), S102--S102.

\bibitem[\protect\citeauthoryear{{The Turing Way Community}}{{The Turing Way
  Community}}{2022}]{the_turing_way_community_2022_6909298}
{The Turing Way Community} (2022, July).
\newblock {The Turing Way: A handbook for reproducible, ethical and
  collaborative research}.

\end{thebibliography}

\end{document}